\documentclass[superscriptaddress, twocolumn]{revtex4-2}
\usepackage{xcolor}
\usepackage{bibunits}
\definecolor{db}{RGB}{46,48,147}
\usepackage{graphicx}
\usepackage{amsmath,amssymb}
\usepackage{dcolumn}
\usepackage{bm}
\usepackage{float}
\usepackage[colorlinks = true,
            linkcolor = db,
            urlcolor  = db,
            citecolor = db,
            anchorcolor = db]{hyperref}
\usepackage{subfiles}

\newcommand{\physrep}{Phys. Rep.}
\newcommand{\jcap}{J. Cosml. Astropart. Phys.}
\newcommand{\araa}{Annu. Rev. Astron. Astrophys.}
\newcommand{\mnras}{Mon. Not. R. Astron. Soc.}
\newcommand{\apjl}{Astrophys. J. Lett.}

\begin{document}

\title{Possible High-Energy Neutrino Emission from Dark Matter Annihilation in the Disrupting Dwarf Galaxy Bo\"{o}tes~III}

\author{Shunhao Ji}
\affiliation{Department of Astronomy, School of Physics and Astronomy, Key Laboratory of Astroparticle Physics of Yunnan Province, Yunnan University, Kunming 650091, China}

\author{Zhongxiang Wang}
\affiliation{Department of Astronomy, School of Physics and Astronomy, Key Laboratory of Astroparticle Physics of Yunnan Province, Yunnan University, Kunming 650091, China}

\begin{abstract}
We report the first search for high-energy neutrino emissions from dark matter
(DM) annihilation in stellar-stream cores.
Motivated by a recent gamma-ray study that proposed these cores
as a new class of indirect DM targets, we analyze three stream cores in
the Northern Hemisphere
using the public ten-year track-like neutrino data released by IceCube. 
Under the $\chi\chi\to\nu\bar{\nu}$ annihilation hypothesis, the most significant
excess among the three targets is found at the position of the nearby
dwarf galaxy Bo\"{o}tes~III,
the core of the Styx stream, with the DM mass of
$m_\chi\simeq10$--$31~{\rm TeV}$. The excess has a post-trial significance of $3.1\sigma$.  
Considering the existing IceCube
dwarf-galaxy limit for the same channel, we obtain a limit on the J-factor
$J_{\rm ann}$, $\log_{10}(J_{\rm ann}/{\rm GeV^2\,cm^{-5}})
\gtrsim 19.1^{+0.3}_{-0.6}$. This limit is broadly
consistent with empirical estimates of $J_{\rm ann}$ for Bo\"{o}tes~III.
The results provide the first candidate target with a possible HE neutrino
signal associated with DM annihilation.
This neutrino excess and the general existence of DM-induced neutrino signals 
from other similar sources will be confirmed with the near-future large 
high-energy neutrino detectors, thus enabling us to probe the nature of 
DM particles.
\end{abstract}

\maketitle

\section{Introduction}
A wide range of astrophysical and cosmological observations require a
dominant dark matter (DM) component, whose fundamental nature remains unknown
\cite{ber+05}. Weakly interacting massive particles (WIMPs), predicted in many
extensions of the Standard Model (SM), are among the most widely studied DM
candidates \cite{ros+18}. Their possible annihilation or decay into SM particles has
motivated extensive indirect searches for secondary messengers such as gamma
rays, cosmic rays, and neutrinos \cite{gas16}.
High-energy (HE) neutrinos are one of promising messengers for searches for 
DM-induced emissions \cite{arg+21}; unlike charged cosmic rays, 
they are not deflected by magnetic fields, and they are much
less attenuated than gamma-rays at high energies or in dense environments.

The IceCube Neutrino Observatory at the South Pole \cite{Aartsen+17} has opened a new window on the TeV--PeV universe by detecting HE
neutrinos of astrophysical origin \cite{Aartsen+13}.
Identifying the sources of these neutrinos,
however, remains a challenge. To date, only a limited number of
associations have been established. These include the blazar
TXS~0506+056, associated with the 2017 neutrino alert at $\sim3\sigma$ 
significance level \cite{txs0506a} and an earlier neutrino
flare with significance of $3.5\sigma$ \cite{txs0506b}, the nearby Seyfert 
galaxies NGC~1068 and
NGC~4151, reported with significances of $4.2\sigma$ and $\sim3\sigma$,
respectively \cite{ngc1068,Ner+24,Abbasi+25}, and diffuse neutrino emission
from the Galactic plane at the $4.5\sigma$ level \cite{gal}. 
The origin of a large fraction of the astrophysical neutrino flux remains 
to be investigated.
Previous studies have investigated DM-induced neutrinos from a variety of potential targets, including the Galactic center \cite{abb+12,alb+20,iov+21,abb25}, 
the Galactic halo \cite{abb+11}, the Sun \cite{kin,abb12sun,aar17sun,abb22sun} and 
the Earth \cite{aar17earth,abb25earth}, galaxy clusters \cite{aar13GC}, 
and dwarf galaxies \cite{aar13GC,guo+23,lv+24,abb25dg}. However, no potential neutrino signal attributable to DM annihilation has been found from any of these targets.

More recently, Ref.~\cite{fer+25} proposed stellar streams originating from dwarf galaxies
as a new class of targets for indirect DM searches. Dwarf galaxies are expected
to be highly DM dominated \cite{mateo98,sim+07}. Although stellar streams can 
lose most of their original DM halos through
tidal stripping, a significant amount of DM may still remain in their 
cores \cite{pen+08,agu+23}. Annihilation within these cores could therefore 
produce detectable gamma-ray or neutrino emissions.
Ref.~\cite{fer+25} has performed a search for gamma-ray emissions from 
these stream cores by analyzing the data from the Large Area Telescope (LAT)
onboard \textit{Fermi}, but has not found significant signals.
In this work, we perform
the first search for neutrino emissions from DM annihilation in 
dwarf-galaxy stream cores by using IceCube's public ten-year track-like dataset.

The targets are from the \textit{golden} sample of
Ref.~\cite{fer+25}, which consists of eight reliable dwarf-galaxy origin streams
chosen according to DM-motivated criteria.
By further requiring the stream cores with declinations
between $-3^\circ$ and $81^\circ$, where the IceCube has
favorable sensitivity for track-like events \cite{ngc1068},
we are left with three stream-core targets: LMS-1, PS1-D, and Styx.
For LMS-1 and PS1-D, we adopt the core positions from Ref.~\cite{fer+25},
where the stream midpoints were used as the assumed core locations in the
absence of an identified progenitor or core counterpart.
The core of the Styx stream is observationally associated with
the disrupting dwarf galaxy Bo\"{o}tes~III \cite{gri+09,car09,car18}. 
We refer to this target as Bo\"{o}tes~III hereafter.

\begin{figure*}[ht!]
    \includegraphics[width=0.49\textwidth]{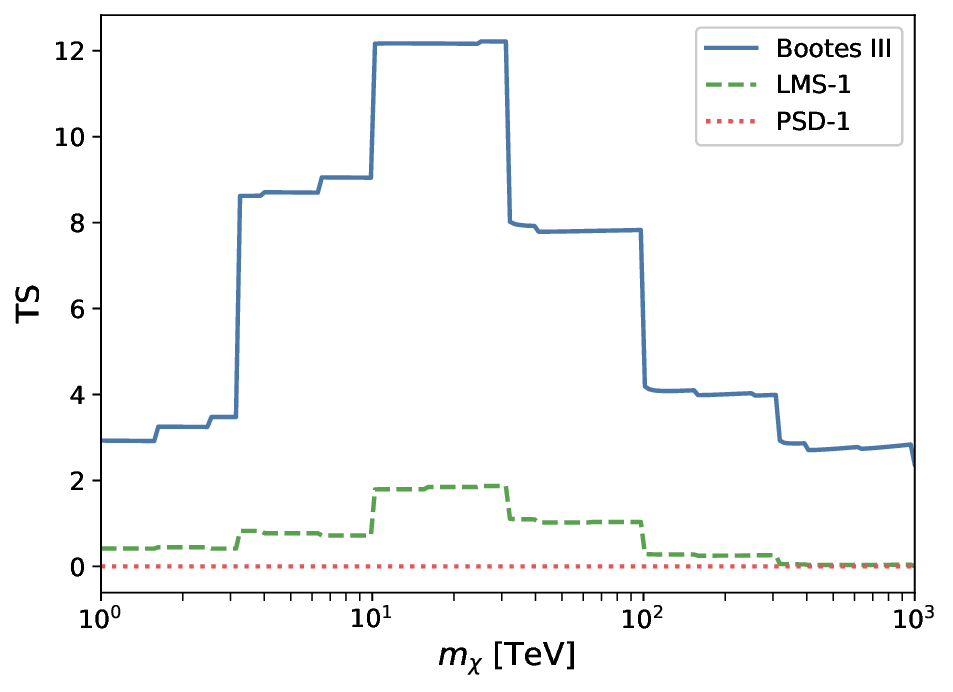}
    \includegraphics[width=0.49\textwidth]{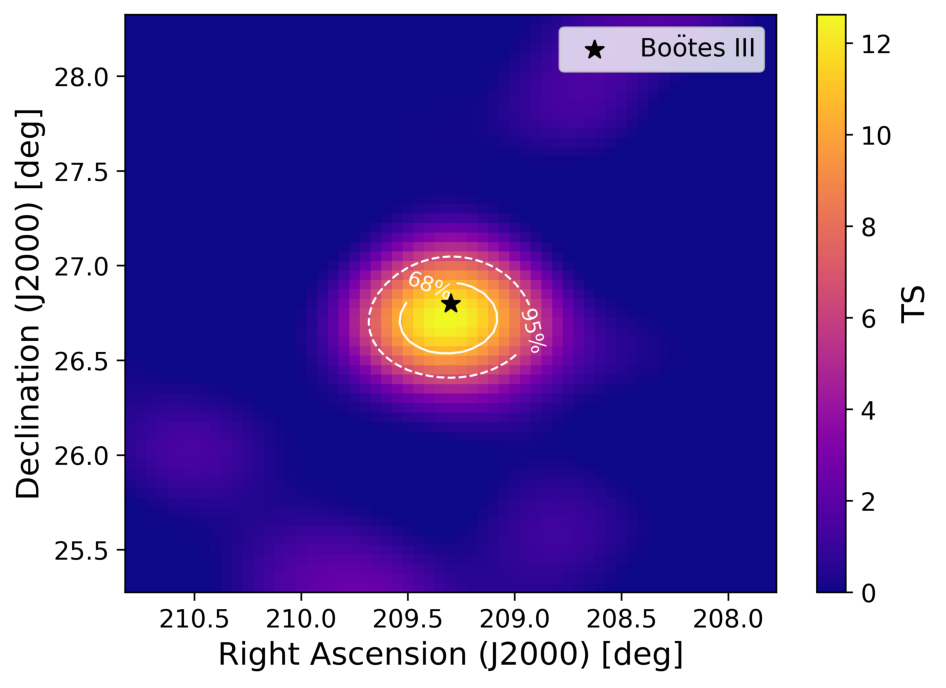}
	\caption{$Left:$ TS values obtained from scanning the DM mass 
$m_\chi$ from 1\,TeV to 1\,PeV for the three selected stream-core targets
	(LMS-1, dashed line; PS1-D, dotted line; Bo\"{o}tes~III, solid line).
The largest TS peak at the best-fit value $m_\chi=26.5\,{\rm TeV}$ is found for
    Bo\"{o}tes~III.  $Right:$ $3^\circ\times3^\circ$ TS map centered at
	the position of Bo\"{o}tes~III (marked with a black star), where 
	the DM mass is fixed at the best-fit value. 
The white solid and dashed contours show the 68\% and 95\% uncertainty regions 
of the neutrino excess spot, respectively. Bo\"{o}tes~III lies within 
the 68\% contour, only $0.1^\circ$ from the TS maximum's location.}
    \label{fig:ts}
\end{figure*}

From analyzing IceCube's neutrino dataset, we find a neutrino excess 
with a post-trial significance of $3.1\sigma$ at the position of 
Bo\"{o}tes~III. This dwarf galaxy, at R.A. = $209.3^\circ$ and 
Dec. = $26.8^\circ$ (J2000), has a distance of $d\simeq46.5$ kpc. It was first 
identified as a diffuse stellar overdensity nearly coincident with the Styx 
stellar stream and was suggested
to be the progenitor of Styx in the final stages of tidal 
disruption \cite{gri+09}. Spectroscopic observations identified candidate 
Bo\"{o}tes~III members with a systemic velocity of 
$V_{\rm helio}=197.5\pm3.8~{\rm km~s^{-1}}$ and a velocity dispersion 
of $14.0\pm3.2~{\rm km~s^{-1}}$, supporting the interpretation of 
Bo\"{o}tes~III as the transitional system between a bound dwarf galaxy and 
an unbound stellar stream \cite{car09}. Gaia proper-motion measurements 
further showed that this dwarf galaxy is dynamically associated 
with Styx and follows a highly eccentric orbit, with a recent pericentric 
passage within $\sim12~{\rm kpc}$ of the Galactic center \cite{car18}.

\section{IceCube Data Analysis and Results}
The public IceCube ten-year dataset contains all-sky track-like events recorded between 
April 2008 and July 2018 \cite{abb21_data}. There were $\sim$1.1 million
events in the dataset, dominated by those of the atmospheric backgrounds. 
Track-like events arise primarily from charged-current
interactions of $\nu_\mu$ ($\bar{\nu}_\mu$) with nucleons and provide an
angular resolution of $\lesssim 1^\circ$ \cite{aar+20}. Thus, they
are the IceCube events of the best class for point-source searches. We use
an unbinned maximum-likelihood method \cite{bra+08} in our
search for neutrino emissions from DM annihilation in the three targets,
and details for the data analysis are provided in the Appendix.

The differential neutrino flux from DM annihilation is given by
\begin{equation}
\frac{d\Phi_\nu}{dE_\nu}
=
\frac{\langle \sigma v \rangle}{8\pi m_\chi^2}
\frac{dN_\nu}{dE_\nu}
\times J_{ann},
\label{dm_dpec}
\end{equation}
where $\langle \sigma v\rangle$ is the velocity-averaged annihilation cross 
section, $m_\chi$ is the DM mass, and $J_{ann}$ is the so-called astrophysical 
$J$-factor, determined by the DM density profile of a target \cite{bod+20}. 
The term $dN_\nu/dE_\nu$ denotes the neutrino spectrum per annihilation for the annihilation channel considered. 
We focus on the most direct channel $\chi\chi\to\nu\bar{\nu}$ \cite{arg+21}. Note that this is an average neutrino channel, i.e., assuming $\nu\bar{\nu}=\frac{1}{3}(\nu_e\bar{\nu}_e+\nu_\mu\bar{\nu}_\mu+\nu_\tau\bar{\nu}_\tau$).
We obtain annihilation spectra for $m_\chi$ from 1\,TeV to 1\,PeV 
using \textsc{HDMSpectra} \cite{Bauer+20}, which include both the continuum 
and line components. After accounting for neutrino oscillations following 
Ref. \cite{abb25dg}, the resulting $\nu_\mu\bar{\nu}_\mu$ spectra at the Earth 
are used as signal templates in the likelihood analysis (see 
Fig.~\ref{fig:spec} in Appendix as an example).
The test statistic (TS) used to evaluate the presence of a signal is defined as
${\rm TS}=2\log\Lambda(n_s,m_\chi)$, where $\log\Lambda$ is the
log-likelihood ratio with respect to the background-only hypothesis and
$n_s$ is the number of signal events for a given $m_\chi$
(see Appendix).

The results from the likelihood analysis of the IceCube data for the three 
stream cores are given in Table~\ref{tab:targets}. The corresponding TS 
profiles obtained from scanning over $m_\chi$ are displayed in the left panel 
of Fig.~\ref{fig:ts}. The largest and only significant excess is at 
the position of Bo\"{o}tes III, with ${\rm TS}=12.2$ at $m_\chi\simeq26.5~{\rm TeV}$.
The corresponding best-fit signal number is $n_s=17^{+8}_{-7}$, where
the uncertainty is at a 68\% confidence level (C.L.) obtained from 
the one-dimensional likelihood scan \cite{Ji+25}. The step-like structure of the TS profile in Fig.~\ref{fig:ts} arises from the binning of the 
publicly available IceCube detector response \cite{abb21_data}, which limits 
the precision of the DM mass determination. We therefore estimate the DM mass,
which lies approximately in the range of $m_\chi\simeq10$--$31~{\rm TeV}$ at a
68\% C.L. for Bo\"{o}tes III.

\begin{table}[t]
\centering
\setlength{\tabcolsep}{6pt}
\begin{tabular}{lccccc}
\hline\hline
Target & R.A.& Dec. & Distance & $n_s$ & TS \\
 & [deg] & [deg] & [kpc] & & \\
\hline
LMS-1              & 231.9 &  27.8 & 18.1 & 7 & 2.1 \\
PS1-D              & 139.7 &  0.8 & 22.9 & 0 & 0.0 \\
Bo\"{o}tes~III/Styx & 209.3 &  26.8 & 46.5 & 17 & 12.2 \\
\hline
\end{tabular}
\caption{
Information for the three stream-core targets and likelihood analysis 
	results.  The TS provides the maximum values obtained in the scan 
over $m_\chi$ under the $\chi\chi\to\nu\bar{\nu}$ annihilation hypothesis
(Note that for LMS-1 and PS1-D, the stream midpoints are adopted as the 
	core positions \cite{fer+25}.)  }
\label{tab:targets}
\end{table}

\begin{figure}[ht]
    \includegraphics[width=0.48\textwidth]{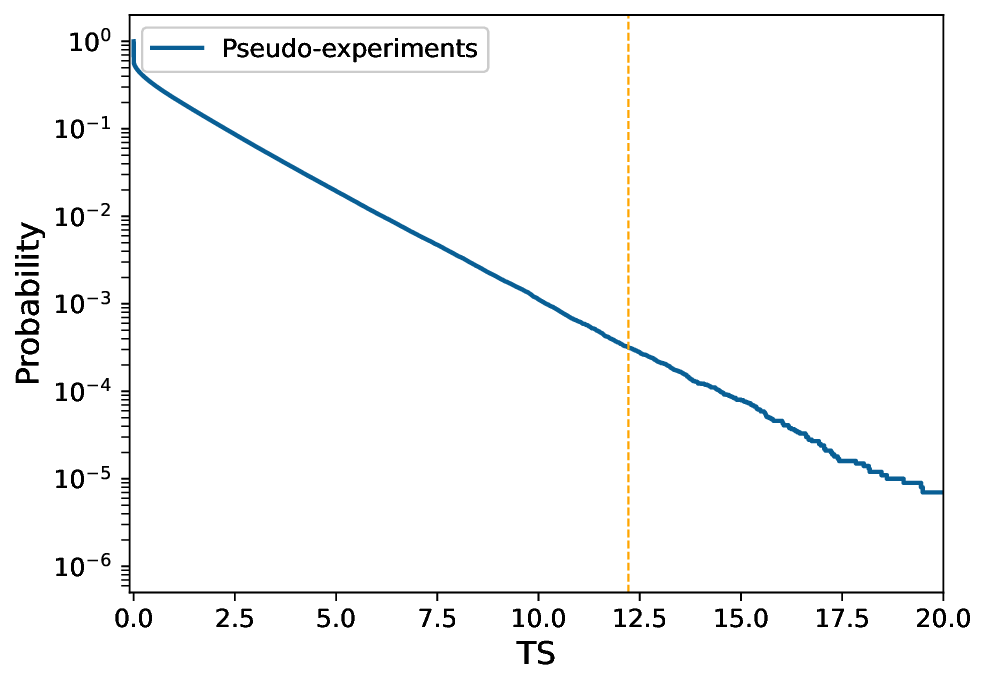}
	\caption{TS distribution obtained from $10^6$ background-only
    pseudo-experiments. The orange dashed line indicates the TS observed at the
    position of Bo\"{o}tes~III in the real data.}
    \label{fig:dis}
\end{figure}

To estimate the significance of the excess at Bo\"{o}tes~III, we 
generate $10^6$ background-only pseudo-experiments and repeat the full scan 
over $m_\chi$ in each trial. This procedure accounts for the look-elsewhere effect associated with the DM-mass scan \cite{gross+10}.
The resulting background-only TS distribution is shown in Fig.~\ref{fig:dis}.
The $p$-value is estimated as the fraction of pseudo experiments whose
maximum TS values exceed that obtained from the real data.
From the TS distribution, a pre-trial $p$-value $p_{\rm pre}$ is found to 
$p_{\rm pre}\simeq3.2\times10^{-4}$, which corresponds to a significance 
of $3.4\sigma$.
There is a trial factor arising from the search from the number 
of targets $n_t = 3$, the post-trial $p$-value 
$p_{\rm post}=1-(1-p_{\rm pre})^3 \simeq9.5\times10^{-4}$, which corresponds
to a significance of $3.1\sigma$.

We further constructed a TS map by scanning a $3^\circ\times3^\circ$ region
around Bo\"{o}tes III with a bin size of $0.05^\circ$ (the right panel of Fig.~\ref{fig:ts}).
The maximum TS=12.6, found at R.A. = $209.3^\circ$ and Dec. = $26.7^\circ$
(J2000), only $0.1^\circ$ away from the position of Bo\"{o}tes~III.  
The 68\% and 95\% uncertainty regions for the neutrino excess spot are 
estimated using Wilks' theorem with two degrees of freedom \cite{Wilks}. 
Bo\"{o}tes~III lies within the 68\% uncertainty region.

\section{Discussion}
By analyzing the public ten-year track-like
neutrino data released by IceCube, we conduct the
search for HE neutrino emissions from DM annihilation in 
three stellar-stream cores in the Northern Hemisphere; this type of the sources
were recently proposed as possible sites of DM annihilation and tested 
with \textit{Fermi}-LAT gamma-ray observations \cite{fer+25}. We
find a neutrino excess spatially coincident with the disrupting dwarf galaxy 
Bo\"{o}tes~III, the core of the Styx stream. Under the DM annihilation signal 
hypothesis in channel $\chi\chi\to\nu\bar{\nu}$, the excess reaches a 
post-trial detection significance of $3.1\sigma$. Although the 
significance is relatively low and the analysis relies on
the certain hypothesis, the result provides the first candidate target with 
a possible HE neutrino signal associated with DM annihilation.

\begin{figure}[h]
    \includegraphics[width=0.49\textwidth]{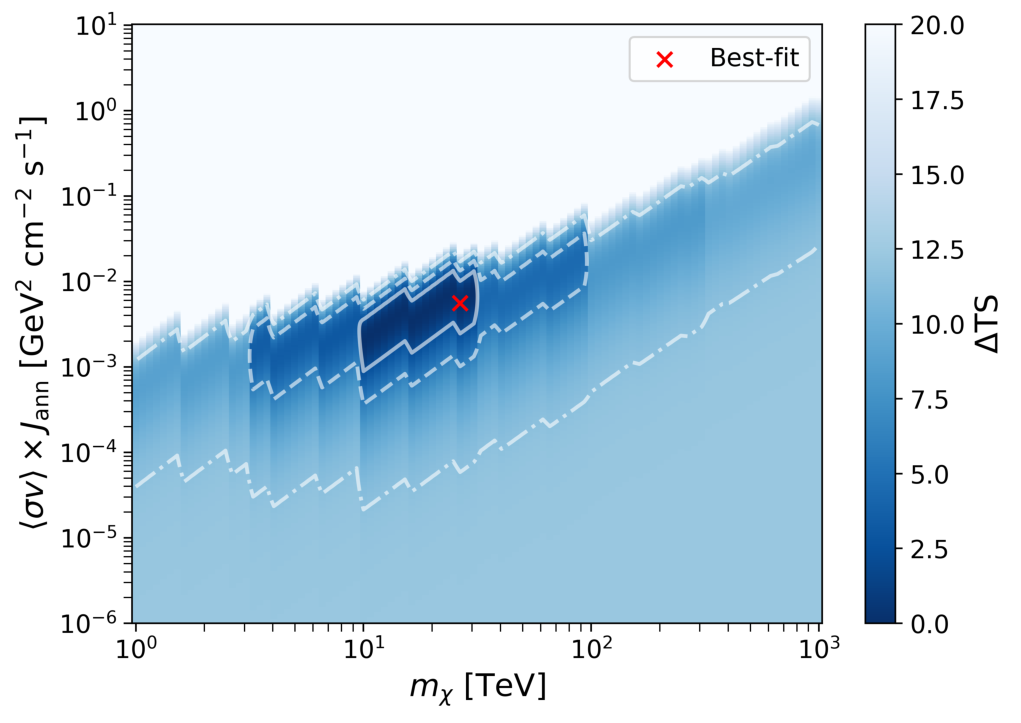}
	\caption{Two-dimensional likelihood scan in the $m_\chi$--$\langle\sigma v\rangle J_{\rm ann}$ plane for the Bo\"{o}tes~III neutrino excess. The red cross marks the best-fit point. The white solid, dashed, and dash-dotted contours show the 68\%, 95\%, and 99.7\% confidence regions, respectively, obtained using Wilks' theorem with two degrees of freedom.}
    \label{fig:llh}
\end{figure}

Considering the DM annihilation as the origin of the neutrino excess,
we explore the corresponding particle parameter space. 
Since Bo\"{o}tes~III is a disrupting dwarf galaxy in transition between
a bound dwarf system and an unbound stellar stream, its J-factor
$J_{\rm ann}$ is substantially more uncertain than that of a dwarf galaxy
in dynamic equilibrium. We do not attempt to determine
$\langle\sigma v\rangle$ and $J_{\rm ann}$ separately, but instead we scan the
two-dimensional likelihood in the plane of $m_\chi$ and
$\langle\sigma v\rangle J_{\rm ann}$, to which the neutrino flux is directly
proportional (see also Ref.~\cite{alex15}).
The resulting allowed parameter space is shown in Fig.~\ref{fig:llh}, 
in which the 68\%, 95\%, and 99.7\% confidence regions are given, estimated from
using Wilks' theorem with two degrees of freedom.
As a reference, IceCube has reported a 90\% C.L. upper limit on 
the annihilation cross section for the $\chi\chi\to\nu\bar{\nu}$ channel 
from dwarf-galaxy searches, 
with  $\langle\sigma v\rangle \simeq 5\times10^{-22}\,{\rm cm^3\,s^{-1}}$ at $m_\chi\simeq26.5\,{\rm TeV}$ (see Fig.~3 of Ref.~\cite{abb25dg}). 
Taking this upper-limit value, the excess we find would require
$\log_{10}({J_{\rm ann}}/{\rm GeV^2\,cm^{-5}})
\gtrsim 19.1^{+0.3}_{-0.6}$,
where the uncertainty corresponds to the 95\% confidence interval. Based on the photometric scaling relations,
$\log_{10}({J_{\rm ann}}/{\rm GeV^2\,cm^{-5}}) =18.65\pm0.60$ was estimated
for Bo\"{o}tes~III \cite{dan+24}, although this estimate is subject to large 
systematic uncertainties due to
the disrupted nature of Bo\"{o}tes~III. We note that the two values
are consistent with each other within the uncertainties.

There are several reasons in favor of the excess detection toward 
Bo\"{o}tes~III. First, it is a well-defined target in the sample
as the core of Styx. For comparison,
the progenitors or compact cores have not been clearly identified 
for LMS-1 and PS1-D, possibly
because either they are nearly completely dissolved or the surviving
remnants have not been located.
Second, its distance ($d\simeq46.5\,{\rm kpc}$) 
places Bo\"{o}tes~III among the relatively
nearby dwarf-galaxy systems used in indirect DM searches. In the sample of
Ref.~\cite{dan+24}, it lies within approximately the nearest 30\% of the
targets. Third, it benefits from its Northern-sky position in 
the IceCube observations, since the downgoing atmospheric muons are
suppressed in the track-like data.
Finally, under the $\chi\chi\to\nu\bar{\nu}$ annihilation hypothesis, the
expected neutrino spectrum is relatively hard 
(see Fig.~\ref{fig:spec} in Appendix).
Such a spectrum is easier to be distinguished from the atmospheric background,
which is dominated by lower-energy events,
and to be detected in IceCube point-source searches \cite{aar+20}.

Except the DM annihilation in dwarf galaxies and their tidal remnants,
no strong HE processes, such as particle acceleration, are expected to occur 
in them.
In order to check the possibility that an unrelated HE source in coincidence
with the direction of Bo\"{o}tes~III,
we have searched for possible counterpart
candidates. Within the 95\% uncertainty region of the neutrino excess spot,
no known types of HE neutrino sources, such as hard X-ray active galactic 
nuclei \cite{ngc1068,Ner+24,Abbasi+25,abb+26} and gamma-ray blazars
\cite{txs0506a,txs0506b}, are found, where the Swift/BAT
all-sky hard X-ray survey catalog \cite{lien25} and the
\textit{Fermi}-LAT source catalog \cite{bal23} are considered.

Future neutrino data will be essential for testing whether the neutrino excess 
persists and for extending the search to additional stream cores. More than 
half of targets in the initial stream-core sample we choose lie in 
the Southern sky, where the IceCube does not have its optimal sensitivity. 
Current and next-generation neutrino observatories in the Northern Hemisphere 
or at mid-latitudes, including KM3NeT \cite{adr16}, Baikal-GVD \cite{bel22}, 
TRIDENT \cite{ye23}, HUNT \cite{Huang23}, and NEON \cite{zhang25}, will 
provide sensitive coverage for the Southern-sky targets.
Improved dynamical modeling of the Bo\"{o}tes~III system is also needed to determine whether a sufficiently dense DM remnant can survive in the stream core and to reduce the uncertainty on the $J$ factor.

\begin{acknowledgments}
This research is supported by the National Natural Science Foundation of
China (12273033) and the Xingdian Talent Support Project of
the Yunnan Province (XDYC-YLXZ-2023-0016).
S. J thanks the support from the Project of Yunnan Provincial Department of Education Science Research Fund (2026Y0169).
\end{acknowledgments}

\bibliographystyle{apsrev4-2}
\bibliography{main}

\appendix*
\section{IceCube Data analysis}
\label{app:icecube}
The public IceCube ten-year point-source dataset \cite{abb21_data} has been
widely used in previous searches for neutrino emissions \cite{aar+20, zhou21, abb22_10,Ner+24,li25,Ji+25}. We grouped the data into
five event samples corresponding to different detector configurations: IC40,
IC59, IC79, IC86\_I, and IC86\_II--VII \cite{aar+20}. An unbinned maximum-likelihood 
method was used to test for a signal from the DM annihilation \cite{bra+08}. 
The likelihood function $\mathcal{L}$ is written as
\begin{equation}
\mathcal{L} =
\prod_{k}
\prod_{i=1}^{N_k}
\left(
\frac{n_s^k}{N_k} S_i^k
+
\frac{N_k - n_s^k}{N_k} B_i^k
\right),
\end{equation}
where $N_k$ and $n_s^k$ are the total number of events and the number of
signal events in sample $k$, respectively. The quantities $S_i^k$ and $B_i^k$
are the signal and background probability density functions (PDFs) for the
$i$th event in sample $k$. 

The signal PDF is divided into the spatial and energy components
\begin{equation}
S_i^k
=
\mathcal{S}_S^k(\vec{x}_i,\sigma_i \mid \vec{x}_s)
\,
\mathcal{E}_S^k(E_i \mid \delta_s, p_s),
\label{eq:signal_pdf}
\end{equation}
where $\mathcal{S}_S^k$ describes the detector's point-spread function (PSF) for
a source at position $\vec{x}_s$. We approximated this term by a 
two-dimensional Gauss \cite{huang+22},
\begin{equation}
\mathcal{S}_S^k(\vec{x}_i,\sigma_i \mid \vec{x}_s)
=
\frac{1}{2\pi\sigma_i^2}
\exp\left(
-\frac{\psi_i^2}{2\sigma_i^2}
\right)\frac{\psi_i}{\sin(\psi_i)},
\label{eq:spatial_pdf}
\end{equation}
where $\psi_i$ is the angular separation between the reconstructed direction
$\vec{x}_i$ for event $i$ and the source position $\vec{x}_s$, and
$\sigma_i$ is the angular uncertainty of the event.

The term $\mathcal{E}_S^k$ is the signal energy PDF, which gives the
probability density for observing a signal event with reconstructed energy
$E_i$ from a source at declination $\delta_s$. It is written as \cite{huang+22}
\begin{equation}
\begin{split}
\mathcal{E}_S^k (E_i \mid \delta_s,p_s)
&= \\[-0.4em]
&\hspace{-0.1cm}
\frac{
\int \Phi_\nu (E_\nu, p_s)
A_{\rm eff}^{k}(E_\nu,\delta_s)\,
\mathcal{M}_k(E_i \mid E_\nu,\delta_s)\,
dE_\nu
}{
\int \Phi_\nu (E_\nu, p_s)
A_{\rm eff}^{k}(E_\nu,\delta_s)\,
dE_\nu
}.
\end{split}
\end{equation}
where $p_s$ denotes the parameter of the assumed source spectrum
$\Phi_\nu (E_\nu, p_s)$. For example, for 
neutrino energy $E_\nu$ of a power-law spectrum,
$\Phi_\nu\propto E_\nu^{-\gamma}$ and $p_s=\gamma$. The quantities
$A_{\rm eff}^{k}$ and $\mathcal{M}_k$ are the effective area and smearing
matrix of the detector for sample $k$, respectively, provided in the public
IceCube data release.

\begin{figure}[ht]
    \includegraphics[width=0.48\textwidth]{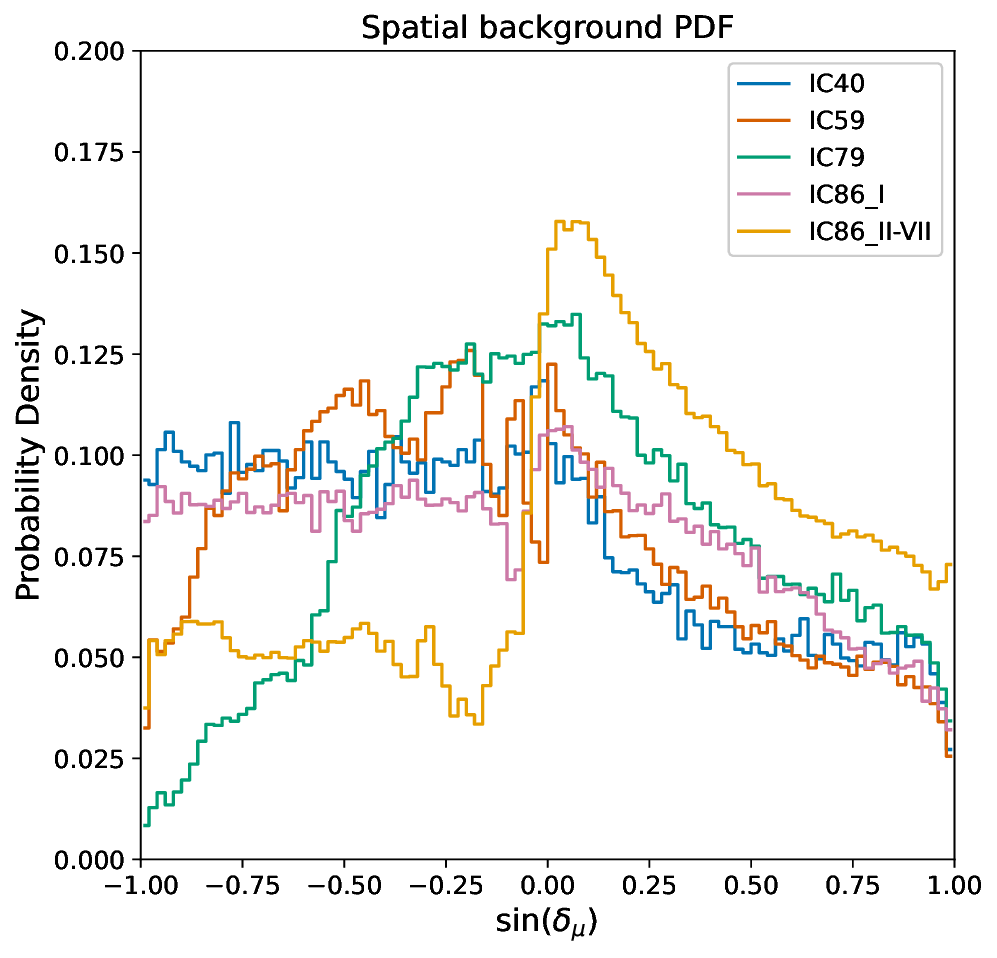}
	\caption{Binned spatial background PDFs for five IceCube data samples
	used in the likelihood analysis (same as Ref.~\cite{huang+22}).}
    \label{fig:bpdfs}
\end{figure}

\begin{figure}[ht]
    \includegraphics[width=0.48\textwidth]{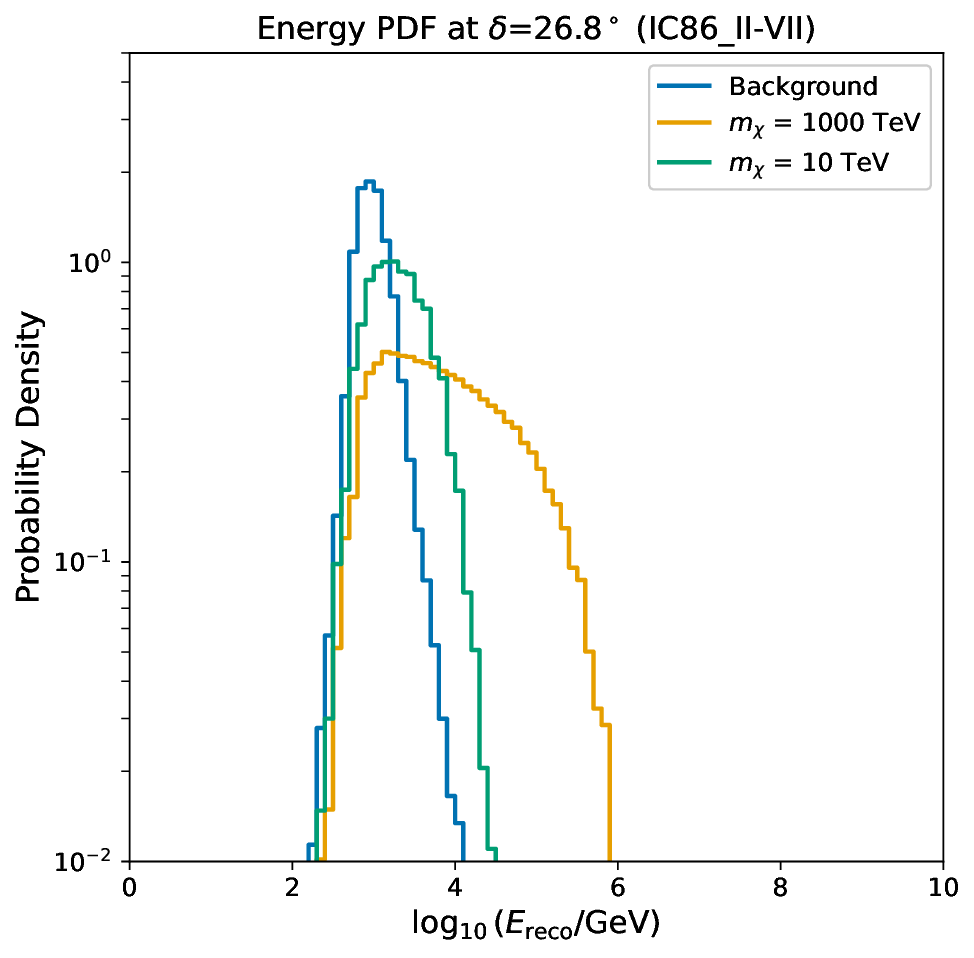}
	\caption{Comparison of the background and signal energy PDFs at the
    Bo\"{o}tes~III declination ($\delta_s=26.8^\circ$), for which 
	the IC86\_II--VII
    sample is taken as an example. The blue curve denotes the background 
	energy PDF, and the green and
    orange curves denote the signal energy PDFs for $m_\chi=10\,{\rm TeV}$ and
    $m_\chi=1000\,{\rm TeV}$, respectively.}
    \label{fig:epdfs}
\end{figure}

The background PDF is also factorized into spatial and energy components,
\begin{equation}
\begin{aligned}
B_i^k
&=
\mathcal{S}_B^k(\delta_i)\,
\mathcal{E}_B^k(E_i,\delta_i)
\\
&=
\frac{1}{2\pi}
P_B^k(\delta_i)\,
\mathcal{E}_B^k(E_i,\delta_i) .
\end{aligned}
\label{eq:bg_pdf}
\end{equation}
The background spatial term depends only on the event declination $\delta_i$
(i.e., the PDF $P_B^k(\delta_i)$),
because the background distribution is uniform in right ascension for IceCube.
The background energy PDF describes the reconstructed energy distribution of
background events and is derived directly from the data. We estimate the background PDFs using bins of $\Delta\sin\delta_\mu=0.02$ in reconstructed declination
and $\Delta\log_{10}E_\mu=0.1$ in reconstructed energy \cite{huang+22}. 
Fig.~\ref{fig:bpdfs} shows the binned spatial PDFs and
Fig.~\ref{fig:epdfs} shows the binned energy PDF from the IC86\_II--VII
sample as an example. 

The relative signal weight of sample $k$ is given by
\begin{equation}
w_k
=
\frac{
t_k
\int \Phi_\nu (E_\nu, p_s)
A_{\rm eff}^{k}(E_\nu,\delta_s)\,
dE_\nu
}{
\sum_{k'} t_{k'}
\int \Phi_\nu (E_\nu, p_s)
A_{\rm eff}^{k'}(E_\nu,\delta_s)\,
dE_\nu
},
\end{equation}
where $t_k$  is the detector livetime for sample $k$, and $k'$ represents 
data samples other than $k$. The expected number of
signal events in sample $k$ is written as $n_s^k=n_s w_k$, where $n_s$
is the total number of signal events.

The log-likelihood ratio with respect to the background-only hypothesis
($n_s=0$) is
\begin{equation}
\hspace*{-0.35cm}
\begin{aligned}
\log \Lambda(n_s,p_s)
&=
\log
\frac{\mathcal{L}(n_s,p_s)}
{\mathcal{L}(n_s=0)}
\\
&=
\sum_k \sum_{i=1}^{N_k}
\log \Bigg[
1+
\frac{n_s w_k}{N_k}
\left(
\frac{S_i^k}{B_i^k}-1
\right)
\Bigg] .
\end{aligned}
\label{eq:loglike_ratio}
\end{equation}
The test statistic (TS) is defined as
\begin{equation}
{\rm TS}=2\log\Lambda(n_s,p_s).
\end{equation}
The best-fit values $n_s$ and $p_s$ are obtained by maximizing the likelihood ratio.

\begin{figure}[h]
    \includegraphics[width=0.48\textwidth]{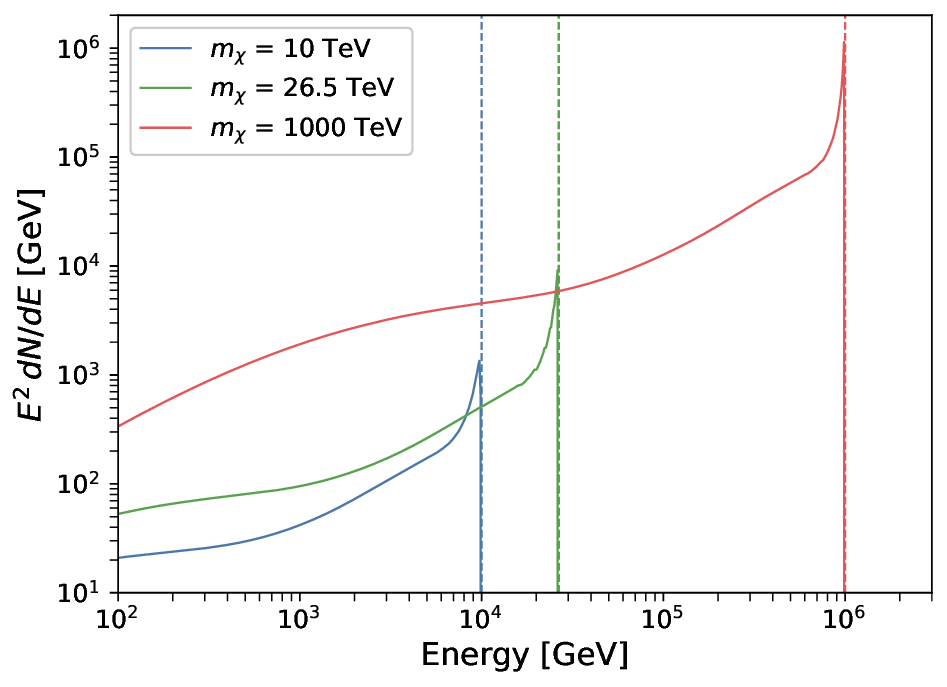}
	\caption{Oscillated $\nu_\mu\bar{\nu}_\mu$ spectra at the Earth for the
    $\chi\chi\to\nu\bar{\nu}$ channel, generated with \textsc{HDMSpectra}. The
    curves show $E^2 dN/dE$ for $m_\chi=10$, 26.5, and 1000\,TeV. The vertical
   dashed lines mark the corresponding DM masses, where the line-like neutrino
   component is located. Both continuum and line components are included in the analysis.}
    \label{fig:spec}
\end{figure}

We validated our analysis framework by testing the data analysis for
the active galaxy NGC~1068, which has been reported as an HE neutrino source 
by the IceCube Collaboration \cite{aar+20,ngc1068}.
In this validation, we assumed a power-law neutrino energy
spectrum, $\Phi_\nu\propto E^{-\gamma}$. Our analysis gave $n_s=52^{+16}_{-14}$, $\gamma=3.2^{+0.4}_{-0.3}$, and flux at 1\,TeV $\Phi_{\rm 1TeV}=3.2^{+1.0}_{-1.0}\times10^{-11}\,{\rm TeV}^{-1}{\rm cm}^{-2}{\rm s}^{-1}$ (all uncertainties are at 68\% C.L.), consistent with those from the previous ten-year IceCube point-source analysis \cite{aar+20}.

For our DM annihilation search, the source spectrum $\Phi_\nu$ was
replaced by the flux model in Eq.~(1) of the main text, for which
$\Phi_\nu\propto \frac{dN_\nu}{dE_\nu}$ for a given $m_\chi$. We obtained
the DM annihilation spectral templates $dN_\nu/dE_\nu$ for the $\chi\chi\to\nu\bar{\nu}$ channel for $m_\chi$ from $1\,{\rm TeV}$ to $1\,{\rm PeV}$ at 100 logarithmically spaced mass points, using \textsc{HDMSpectra} \cite{Bauer+20}.
Examples of the annihilation spectra for the $\nu_\mu\bar{\nu}_\mu$ flavor
at the Earth after considering neutrino oscillations are shown in 
Fig.~\ref{fig:spec}.
The likelihood analysis was then performed for each $m_\chi$ template and 
for each target source.

We estimated the significance of the neutrino excess by performing Monte Carlo 
simulations of background-only pseudo-experiments. The pseudo-data were 
generated by randomizing right-ascension value of each event uniformly in 
the range [0, $2\pi$) while keeping other parameters (declination, reconstructed energy, angular uncertainty,
and data sample) unchanged. This procedure exploits the approximate uniformity 
of the IceCube background in right ascension.
For each pseudo-experiment, we repeated the same analysis procedure as 
applied to the real data, including the scan over $m_\chi$. Therefore,
the look-elsewhere effect associated with the choice of the DM
mass \cite{gross+10} was taken into account.
We generated $10^6$ background-only pseudo-experiments. The $p$-value was 
estimated as
the fraction of pseudo-experiments whose maximum TS values exceeded the value 
obtained from the real data. The observed maximum TS for Bo\"{o}tes~III was 
${\rm TS}=12.2$, estimated to have $p_{\rm pre}\simeq3.2\times10^{-4}$
(corresponding to a significance of $3.4\sigma$).
After applying a trials factor of three for the number of the stream cores 
tested in this work, the post-trial $p$-value is
$p_{\rm post}\simeq9.5\times10^{-4}$ (corresponding to $3.1\sigma$).
\end{document}